\documentclass[aps,prb,twocolumn,superscriptaddress,showpacs]{revtex4}
\usepackage{graphicx,epsfig}
\usepackage{amssymb,times}
\usepackage{amsmath,amsfonts,bm}

\newcommand \dagga {{\phantom{\dagger}}}
\newcommand{\vect}[1]{\mathbf{#1}}

\begin{document}

\title{Cellular Dynamical Mean Field Theory of the Periodic Anderson Model}
\author{Lorenzo De Leo}
\affiliation{Department of Physics and Center for Material Theory,
Rutgers University, Piscataway, NJ 08854, USA}
\author{Marcello Civelli}
\affiliation{Institut Laue Langevin, 6 rue Jules Horowitz - 38042
Grenoble Cedex, France}
\author{Gabriel Kotliar}
\affiliation{Department of Physics and Center for Material Theory,
Rutgers University, Piscataway, NJ 08854, USA}
\begin{abstract}
We develop a cluster dynamical mean field theory of the periodic Anderson
model in three dimensions, taking a cluster of two sites as a
basic reference frame. The mean field theory displays the basic
features of the Doniach phase diagram: a paramagnetic Fermi liquid
state, an antiferromagnetic state and a transition between them.

In contrast with spin density wave theories, the transition is
accompanied by a large increase of the effective mass everywhere
on the Fermi surface and a substantial change of the Fermi surface
shape across the transition. To understand the nature and the
origin of the phases near the transition, we investigate the
paramagnetic solution underlying the antiferromagnetic state, and
identify the transition as a point where the $f$ electrons decouple
from the conduction electrons undergoing an orbitally selective Mott
transition. This point turns out to be
intimately related to the two impurity Kondo model quantum
critical point. In this regime, non local correlations become
important and result in significant changes in the photoemission
spectra and the de Haas-van Alphen frequencies.
The transition involves considerable $f$ spectral weight transfer
from the Fermi level to its immediate vicinity, rather than to the
Hubbard bands as in single site DMFT.
\end{abstract}

\maketitle

\section{Introduction}

The heavy fermion materials have been one of the most challenging
problems for the condensed matter community in the last three
decades. The physics of these systems is dominated by the
competition of many interactions, manifesting in a variety of
phases\cite{Stewart01}. Close to the zero-temperature
transition between two such phases the materials typically show
non-Fermi liquid behavior on a large temperature range. This
behavior has been often attributed to an underlying quantum
critical point (QCP).

In a subclass of these materials (including CeCu$_{6-x}$Au$_x$,
YbRh$_2$Si$_2$ and Ce$T$In$_5$, with $T$=Ir, Co, Rh) it is
possible to drive the system from an antiferromagnetic (AF) phase
to a paramagnetic (PM) phase by suppressing continuously the
N\'{e}el temperature via application of pressure, magnetic field
or doping. The conventional spin-density wave (SDW) theory of a
QCP\cite{Hertz75, Millis93, Moriya95} disagrees with the
experimental evidence in many of these
materials\cite{Stewart01,Schroeder00,Gegenwart02,Custers03,Coleman01}.
This has triggered a renewed interest in microscopic descriptions
of the heavy fermion state and its possible instabilities to
ordered states.
More generally, not only the critical point but also the formation
of the heavy fermion state in the regime where the RKKY interactions
are sizeable requires nontrivial techniques for its microscopic description
\cite{Nakatsuji04,Sun05b}.
Heavy fermion materials can be
modeled by the periodic Anderson model which describes a wide
conduction band of $spd$ electrons and a narrow band of $f$
electrons which is strongly correlated and hybridized with the
$spd$ electrons.

Doniach\cite{Doniach77} established that for large hybridization
the $f$ electrons combine with the conduction electrons to form a
heavy Fermi liquid, while for small hybridization the $f$ moments
order magnetically and hence do not undergo the Kondo effect.
Thus, changes in the hybridization not only lead to the appearance
of magnetic long range order but introduce also a change in the
electronic structure referred to as $f-c$ decoupling or Kondo
breakdown (KB). Namely, for large values of the
hybridization the quasiparticles are composites of $f$ electrons
and conduction $c$ electrons with a Fermi surface which obeys
Luttinger theorem, counting both the $f$ and $c$ electrons. For
small hybridization the $f$ electrons decouple from the conduction
electron quasiparticles, and the Fermi surface contains only the
latter. The decoupling between the $f$ and $c$ electrons can be
viewed as an orbitally selective Mott transition, and has been
studied
with single site DMFT\cite{demedici05a}
and slave boson techniques\cite{pepin-2006}.
As stressed in \cite{Senthil04}, a precise formulation of KB
requires the elimination of antiferromagnetic long range
order which breaks the translational symmetry of the original
lattice. So far this has been achieved by studying models with
disorder within DMFT\cite{Burdin02}, or by modifying the symmetry
group of the spin from SU(2) to SU(N)\cite{Senthil04}.
For earlier studies of the KB using slave bosons see
\cite{Andrei89,Castellani91}.

Single site Extended DMFT of various models have been carried out.
For the Anderson lattice model\cite{Sun03,Sun05a}, the KB
and the onset of AF should be viewed as distinct
instabilities which occur at two different values of the RKKY
coupling (relative to the Kondo
coupling). The same situation occurs in the field theory approach
of Ref. \cite{Senthil04}.

On the other hand, extended DMFT studies of a Kondo lattice model,
in which the conduction electron band is constrained so as to
prevent its magnetic polarization, reveal that the KB
and the onset of magnetism take place at the same point in the
phase diagram\cite{Si01,Grempel03,Zhu03,zhu-2006}.
See however Ref.\cite{glossop-2006}. Justification for the study
of this model is given in Ref.\cite{Si05}.

In DMFT based approaches the breakdown of Fermi liquid theory at
the transition is understood in terms of local physics.
In other works the same breakdown is explained in terms of the
coupling of the electrons to the long wave length fluctuations of
an emergent gauge field\cite{Senthil04} and to the Kondo order
parameter fluctuations \cite{Paul07,pepin-2006}.

In this paper we handle the competition between magnetism and formation of a
heavy PM state in an unbiased way using Cellular Dynamical
Mean Field Theory (CDMFT)\cite{Kotliar2001spb}. This approach
improves the treatment of non-local correlations over single-site
DMFT considering a cluster of sites and providing a
self-consistency condition on the cluster action. All the
correlations whose range is contained within the cluster are then
treated beyond a simple mean-field level. In this paper we
consider a cluster of two sites embedded in a three dimensional
lattice, which is the minimal unit able to capture the competition
between PM and AF correlations. A similar approach was introduced
in an earlier publication\cite{Sun05b} based on a quantum Monte
Carlo impurity solver, but that study was limited to large
temperatures and small values of $U$. We overcome this limitation
by using Exact Diagonalization (ED) as an impurity
solver\cite{Caffarel94}. This solver gives access to the ground
state properties of the system with a finite energy resolution
(typically 0.005 of the bandwidth in our case).

In the theory of the Mott transition, the investigation of the PM
insulating phase has been very fruitful and allowed an
understanding of the finite temperature state, above the magnetic
order in many compounds. We pursue a similar approach to the
periodic Anderson model by investigating both the magnetic and a
``parent compound'' PM state, to understand better the character
of the electronic structure. This allows us to describe the
magnetic state as a result of an additional instability of the PM
state.

We show that a phase transition separates the heavy fermion
paramagnetic phase from an antiferromagnetic phase. Approaching
the transition from the paramagnetic side the Fermi liquid is
strongly renormalized. An analysis of the mean field PM solution
into the AF phase reveals an orbitally selective Mott decoupling
and localization of the $f$-electrons. Within the resolution of
our method, the antiferromagnetic transition and the decoupling of
the $f$ and $c$ electrons take place for the same value of the
hybridization. Through a comparison with a Wilson's Numerical
Renormalization Group (NRG) calculation, we further relate the
nature of the transition to the proximity of the system in the
parameter space to the quantum critical point of the two-impurity
Kondo model (2IKM)\cite{Jones88,Affleck95}.

The transition has dramatic consequences on the quasiparticle
spectra and on the topology of the Fermi surface which can be
detected by photoemission experiments and by Hall effect and de
Haas-van Alphen frequencies measures.

The paper is organized as follows. In Sec. \ref{methods} we
introduce the model Hamiltonian and the CDMFT method employed. In
Sec. \ref{results} we analyze the results of the CDMFT
calculations and compare them with single site DMFT results and
Numerical Renormalization Group results on the underlying impurity
problem.
Conclusion are presented in Sec. \ref{conclusions}.

\section{Methods \label{methods}}

The three-dimensional periodic Anderson model Hamiltonian is
\begin{eqnarray}\label{Hamiltonian}
    H \!\!&=& \!- t \sum_{\langle ij \rangle}
    c^\dagger_{i\sigma} c^\dagga_{j\sigma}
    - \mu \sum_i c^\dagger_{i\sigma} c^\dagga_{i\sigma}
    + V \sum_i \left( f^\dagger_{i\sigma} c^\dagga_{i\sigma}
    + h.c. \right) \nonumber\\
    &&+ (E_f-\mu) \sum_i f^\dagger_{i\sigma} f^\dagga_{i\sigma}
    + U \sum_i n^f_{i\uparrow} n^f_{i\downarrow}
\end{eqnarray}
where $c^\dagger_{i\sigma}$ creates a conduction band electron at
site $i$ with spin $\sigma$ and $f^\dagger_{i\sigma}$ does the
same for the localized orbitals. The conduction electron bandwidth
$W = 12 |t| = 2$ sets the energy unit.

In the large $U/V$ limit the model maps onto the Kondo lattice
model, where each $f$ orbital is replaced by a spin $\frac{1}{2}$
which is Kondo-coupled to the local conduction electrons with a
coupling $J_K = ( \frac{1}{|E_f-\mu|} + \frac{1}{|U+E_f-\mu|}
)V^2$, where $\rho$ is the conduction electron density of states
at the chemical potential.
It is also possible to define two energy scales in the
problem. The first is the single-impurity Kondo temperature $T_0 =
\exp(-1/2 \rho J_K)$. It is not known a priori how the ``lattice''
Kondo energy is related to $T_0$. The second scale is set by the
magnetic interactions. These are not contained explicitly in this
Hamiltonian, but an effective RKKY coupling is generated in fourth
order perturbation theory in $V$ (second order in the Kondo
coupling, $J_{RKKY} \sim J_K^2/W$). The RKKY coupling oscillates
in real space with a period that is related to the bare conduction
band Fermi momentum, ranging from antiferromagnetic at half
filling to ferromagnetic in the limit of an empty band. A
comparison of these two energy scales allows to understand the
qualitative features of the phase diagram of the
model\cite{Doniach77}. The ratio between Kondo temperature and
RKKY interaction is an increasing function of $V$, hence for small
$V$ we will find a phase in which the magnetic fluctuations
dominate giving rise to an ordered state, while for large $V$ the
Kondo effect dominates and the $f$ moments are screened by the
conduction electrons.

At half filling the Kondo screening of the moments by the
conduction electrons results in an insulator with a gap induced
by the hybridization. In this paper we are interested in metallic
heavy fermions, and we restrict ourselves to study the system away
from half filling.

In the path integral formalism the conduction band can be
integrated out in the action and replaced by an effective
long-range retarded hopping for $f$ electrons. After these
manipulations the Lagrangian of the system is
\begin{eqnarray}\label{Lagrangian}
    {\cal L} &=& \sum_i f^\dagger_{i\sigma}(\tau) \left( \partial_\tau -\mu
    + E_f \right) f^\dagga_{i\sigma}(\tau) \\
    &&+ \sum_{ij} f^\dagger_{i\sigma}(\tau) \Delta^c_{ij}(\tau-\tau')
    f^\dagga_{j\sigma}(\tau')
    + U \sum_i n^f_{i\uparrow} n^f_{i\downarrow} \nonumber
\end{eqnarray}
where the effect of the conduction band is absorbed in the Weiss
field
\begin{equation}\label{Delta-cond}
    \Delta^c(i\omega,\vect{q}) = \frac{V^2}{i\omega+\mu-\varepsilon_{\vect{q}}}
\end{equation}
and $\varepsilon_{\vect{q}} = -2t \sum_{\alpha=x,y,z}
\cos(q_\alpha)$.

In CDMFT the original three-dimensional lattice is now tiled with
a superlattice of clusters and an effective Anderson impurity
action is derived for a single cluster. Finally a self-consistency
condition relates the cluster Green's function to the local
Green's function of the superlattice. The self-consistency is more
easily written as a matrix form
\begin{equation}
   G_{loc}(i\omega) = \sum_{\vec{k}} \left[
   \left(
   \begin{array}{cc}
      i\omega +\mu -E_f & 0 \\
      0 & i\omega +\mu -E_f
   \end{array} \right) \right.
\end{equation}
\begin{displaymath}
   \left. - \frac{V^2}{(i\omega+\mu)^2-\varepsilon_{\vec{k}}^2}
   \left(
   \begin{array}{cc}
      i\omega +\mu & e^{ik_x} \varepsilon_{\vec{k}} \\
      e^{-ik_x} \varepsilon_{\vec{k}} & i\omega +\mu
   \end{array} \right) -
   \left(
   \begin{array}{cc}
      \Sigma_{11}(i\omega) & \Sigma_{12}(i\omega) \\
      \Sigma_{21}(i\omega) & \Sigma_{22}(i\omega)
   \end{array} \right)
   \right]^{-1}
\end{displaymath}
where spin index has been suppressed to simplify the notation. It
is very important to notice that in the large $U/V$ limit the
impurity action becomes equivalent to the action of a two impurity
Kondo model away from particle-hole (p-h) symmetry. At p-h
symmetry this model is known to have a QCP separating a Kondo screened
phase from a phase in which the impurities are decoupled from the
conduction band\cite{Jones88,Affleck95}. Away from p-h symmetry
the QCP is not accessible but still influences the physics in an
intermediate energy range\cite{DeLeo04f,varma-2006}.

In order to solve the cluster-impurity problem we employ ED. In
this method the cluster action is recast in the form of an
Anderson impurity Hamiltonian where the continuous degrees of
freedom of the free electron bath are parametrized by a finite
number $N_b$ of sites:
\begin{eqnarray}\label{H_cluster_ED}
    H_{ED} &=& \sum_{i=1,2} f^\dagger_{i\sigma} \left( E_f - \mu
    \right) f^\dagga_{i\sigma} + U \sum_{i=1,2} n^f_{i\uparrow}
    n^f_{i\downarrow} \\
    && + \sum_{l=1,N_b} \varepsilon_{l\sigma} a^\dagger_{l\sigma}
    a^\dagga_{l\sigma}
    + \sum_{i,l} \left( V_{il\sigma} a^\dagger_{l\sigma}
    f^\dagga_{i\sigma} + h.c. \right). \nonumber
\end{eqnarray}

In this representation the hybridization-function
$\Delta_{\mu\nu}= \mathcal{G}^0_{\mu\nu}-\left(i\omega + \mu
- E_f \right)\delta_{\mu\nu}$, related to
the non interacting impurity Green's function $\mathcal{G}^0_{\mu\nu}=
- \langle f^\dagga_{\mu}f^{\dagger}_{\nu} \rangle_{U=0}$ of the cluster
impurity problem, becomes
\begin{equation}\label{hybrid_cluster}
    \Delta_{ij\sigma}(i\omega) = \sum_{l=1,N_b} \frac{ V^*_{jl\sigma}
    V_{il\sigma} }{i\omega - \varepsilon_{l\sigma}}
\end{equation}
where the indices $i,j=1,2$ represent the cluster sites. In the
figures presented in this work we used 8 sites in the bath, but
the results have been tested in few points using 10 sites.

The ground state and Green's functions of this Hamiltonian are
determined via the Lanczos procedure and the self-consistency
equation in turns allows to derive a new set of bath parameters.
The self-consistency condition is used to determine the $\Delta$
function on a finite mesh of points $\omega_n=(2n+1)\pi/\beta$.
This implies a finite resolution in frequency. We used
$\beta=100$. The process is iterated until convergence is reached.
To study the interplay of antiferromagnetism and Kondo screening
we restrict the study to the metallic regime close to half-filling
where the RKKY interaction is predominantly antiferromagnetic.
This is achieved by fixing the values $U=10$, $E_f=-5.5$ and
$\mu=0.2$ throughout the paper.
Being a mean field theory, DMFT gives the possibility to
selectively allow some instabilities and forbid others. Thus we
will allow the development of AF order and study the realistic
material phase diagram. However we will also force the system to
remain in a PM state to study the ``normal state'' underlying the
AF phase. This will prove important to study the KB
point and to understand the origin of the
transitions in the system.

\section{Results \label{results}}

\begin{figure}
  \begin{center}
    \includegraphics[width=8cm]{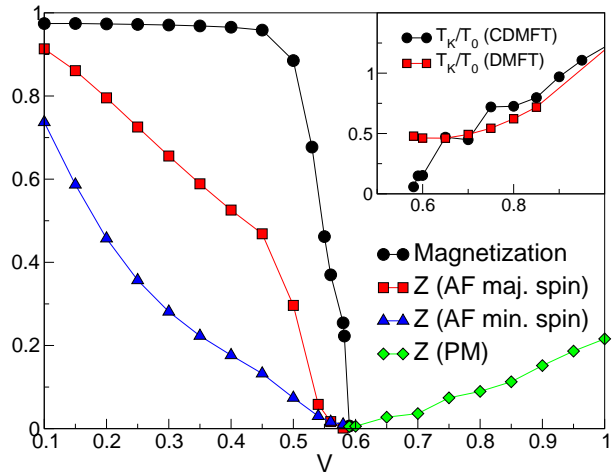}
    \caption{(Color online) Magnetization $M$ and average of the
    quasiparticle residue $Z$ eigenvalues for the AF
    (both for majority and minority spin) and PM solutions
    as a function of $V$.
    Inset: Ratio between lattice and
    single-impurity Kondo energies in DMFT and CDMFT.
    \label{magn}}
  \end{center}
\end{figure}

We compute the
staggered magnetization $M=\langle n^f_{1\uparrow} \rangle -
\langle n^f_{1\downarrow} \rangle = - ( \langle n^f_{2\uparrow}
\rangle - \langle n^f_{2\downarrow} \rangle)$ as a function of $V$
(Fig.\ref{magn}).
For small $V$ the RKKY interaction is dominating and the solution
describes an AF state with a magnetization close to one. In the
opposite limit of large $V$, where the Kondo screening is more
effective, the system is a PM metal. Increasing $V$ in the AF
phase, the magnetization decreases smoothly and approaches zero at
a value $V^* \sim 0.585$ where a phase transition occurs. Above
this critical value the system becomes paramagnetic. The value of
$M$ at the transition jumps from $\sim 0.22$ for $V=0.58$ to $\sim
0.006$ for $V=0.59$ and in between the convergence becomes
extremely slow. The numerical accuracy
does not allow us to exclude a weakly first order transition
scenario, although in this case the
coexistence region would be very small.

To clarify the nature of the transition we investigate the
cluster quasiparticle residue
$\hat{Z}=(\hat{1}-\partial\hat{\Sigma}/\partial\omega)^{-1}|_{\omega=0}$ (in this
case, a matrix in spin and cluster indices). The AF
three-dimensional SDW theory predicts that antiferromagnetism
develops from a well defined Fermi liquid state and hence a finite
quasiparticle residue
$Z$ at the transition. On the contrary KB scenarios
predict
that $Z$ goes to zero at the transition.
In Fig. \ref{magn} we plot the $\hat{Z}$ eigenvalues.
In the AF phase the two curves correspond to minority and majority
species.
In the PM phase we plot only the average of the eigenvalues
since the difference is not appreciable on the scale of the plot.
In both phases the eigenvalues of the $\hat{Z}$ matrix
decrease as we approach the transition.
Precisely at the transition one eigenvalue goes to zero,
suggesting that the KB is indeed the correct
scenario in this model.

Approaching the transition from the PM phase by reducing $V$,
we observe a strong reduction of the $\hat{Z}$ eigenvalues leading to a large effective
mass enhancement.
This result suggests that increasing the correlations
by decreasing $V$ from the PM side, the Fermi liquid becomes
heavier and heavier, and eventually it is more favorable to get
rid of the entropy by developing AF long range order. The
fundamental difference with the SDW picture is that here the
mechanism responsible for the instability is short-ranged in
space.
The cluster self-energies are in fact approximately local
on the PM side of the transition. This may lead to think that the
results are essentially similar to a single site DMFT solution of the
periodic Anderson model. However there are essential differences.
Indeed, even setting the off-diagonal self-energy
$\Sigma_{12}=0$ the CDMFT self-consistency equations do not reduce
to the corresponding single-site DMFT equations.
To demonstrate this
we compare in the inset of Fig.\ref{magn} the ratio between the
lattice Kondo temperature obtained through the relation $T_K =
\Delta(0) Z$ and the corresponding single-impurity energy $T_0$ in
the two cases. While the DMFT result gives a smooth and finite
$T_K$ across $V^*$, in CDMFT $T_K$ appears to approach a zero
value at the transition.

We take advantage now of the possibility of suppressing the AF
long range order to study the normal PM solution in the AF phase.
For $V<V^*$ this solution
is obviously unstable to antiferromagnetism and it corresponds
physically to a highly frustrated system.
\begin{figure}
  \begin{center}
    \includegraphics[width=8cm]{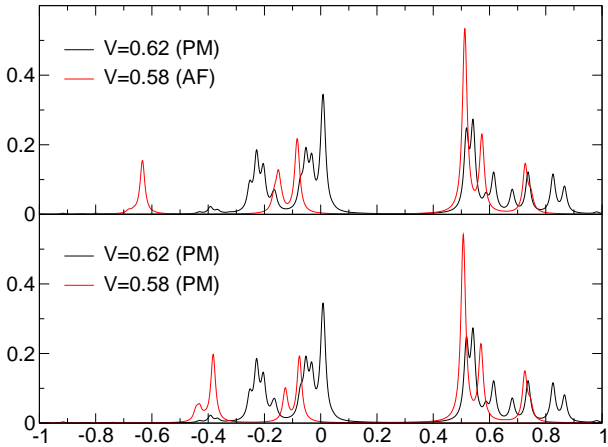}
    \caption{(Color online) Spectral function of the $f$ electrons
    on the two sides of the transition with
    AF order allowed (upper panel) and forbidden (lower
    panel). The two Hubbard bands are outside the energy range.
    \label{spectral_function}}
  \end{center}
\end{figure}
The analysis of the spectral function along the two solutions
(allowing and forbidding magnetic ordering,
Fig.\ref{spectral_function}) reveals that the transition takes
place also in absence of antiferromagnetism.
Namely, at small V, a phase with no spectral weight at the Fermi
level is realized in both cases.
The Kondo peak present in the PM phase disappears in going across the
transition, leaving behind a pseudogap, and the $f$ electrons
effectively decouple from the conduction band, at least within the
energy resolution of our calculations.
The transition is an orbitally selective Mott transition
for the $f$ band.
This implies that the $f$
electrons do not contribute to the Fermi surface for $V<V^*$.
This is in marked contrast with the paramagnetic single site DMFT
solution where there is only a Fermi liquid Kondo screened phase
and no sign of such a transition or of a pseudogapped phase.
Notice also that in the CDMFT treatment the $f$ transfer of spectral
weight takes place from the Fermi level to its immediate vicinity
as the hybridization is reduced. This should be contrasted with
the single site DMFT description of the Mott transition in the Hubbard model where the
spectral weight is transferred from the Fermi level directly to the
Hubbard bands at much higher energies.

\begin{figure}
  \begin{center}
    \includegraphics[width=8cm]{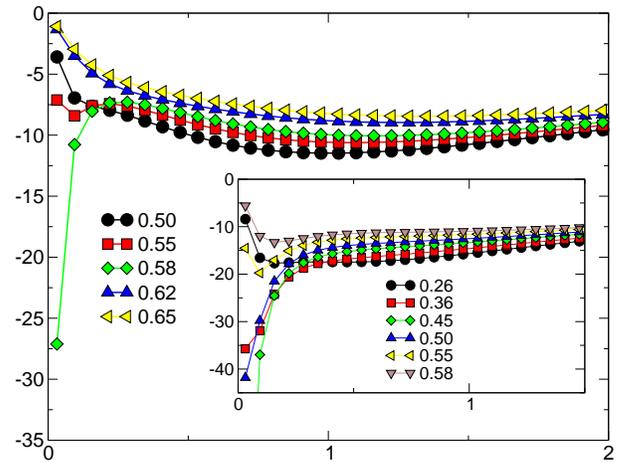}
    \caption{(Color online) Imaginary part of the even self-energy
    eigenvalue on the imaginary axis for different values of the
    hybridization. Inset: odd eigenvalue (in a
    different range of hybridization values).
    \label{sigma_even_odd}}
  \end{center}
\end{figure}
To clarify the Mott character of the transition we plot in
Fig.\ref{sigma_even_odd} the imaginary frequency behavior of the
two $\Sigma$-eigenvalues within the PM solution, named
respectively $\Sigma_{even}=\Sigma_{11}+\Sigma_{12}$ and
$\Sigma_{odd}=\Sigma_{11}-\Sigma_{12}$. Approaching the transition
from the PM phase the slope at zero frequency of both eigenvalues
increases, signaling the growth of the effective mass in the Fermi
liquid. At the transition the even eigenvalue develops a
divergence signaling the end of the Fermi liquid description and
the beginning of the Mott insulating phase. The odd eigenvalue has
a similar divergence at a lower value of the hybridization as
shown in the inset. A qualitative understanding of this behavior
can be proposed considering that $\Sigma_{even}$ and
$\Sigma_{odd}$ represent in a loose sense the lattice self-energy
calculated around momenta $(0,0,0)$ and $(\pi,\pi,\pi)$
respectively. Hence the Mott transition corresponds to the first
appearance of a surface poles in the self-energy around the origin
of the Brillouin zone. Decreasing further the hybridization the
surface of poles moves towards the corner of the zone and finally
disappears for $V\sim0.43$.

\begin{figure}
  \begin{center}
    \includegraphics[width=8cm]{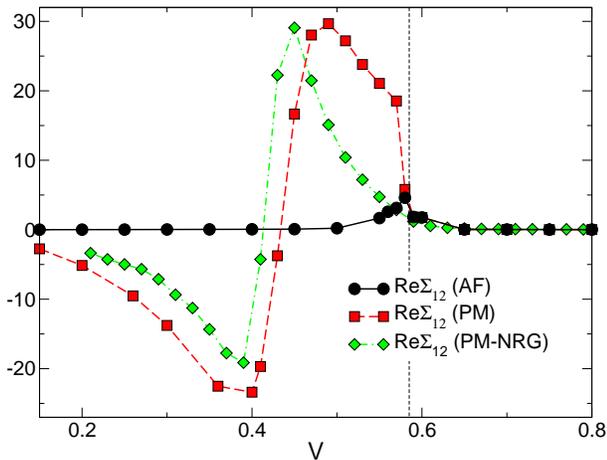}
    \caption{(Color online) Non local correlations (${\rm Re}
    \Sigma_{12}$, calculated at the lowest available imaginary frequency)
    along the AF and PM solutions. The vertical
    dashed line marks the transition. For comparison
    with the PM case the NRG results (without self-consistency).
    \label{sigma12}}
  \end{center}
\end{figure}

While for $V > V^*$ the self-energy was approximately local,
in the PM solution the development of magnetism for $V < V^*$ is
replaced by a dramatic increase of non-local correlations as shown
in Fig.\ref{sigma12}, where we plot the real part of $\Sigma_{12}$
at the lowest available frequency in the imaginary axis.

The behavior of $T_K$, of the spectral function and of the
off-diagonal self-energy near the transition is very similar to
the corresponding behavior found in the 2IKM close to its
QCP\cite{DeLeo04f,varma-2006,in_prep}. To establish this
connection we used Wilson's Numerical Renormalization Group
to solve the impurity model underlying the self-consistency in the
PM case for the same set of parameters.
The difference with the ED calculation is that we used a constant
density of states for the NRG bath, so the results are non
self-consistent.
The results are nevertheless
qualitatively similar. The main difference between the results of
the two calculations is that the transition found in CDMFT is
replaced in NRG by a crossover. We ascribe this effect to the lack
of a self-consistent bath in the NRG calculation. In Fig.\ref{sigma12}
we compare the non local correlations, finding a very good
agreement. Other features like the pseudogap formation around
$V^*$ are also
reproduced. The similarity of these two results suggests that most
of the physics relevant to the lattice is already captured at the
level of the two impurity problem. In this case the onset of
magnetism in the lattice model can be understood as arising from
the vicinity to the impurity QCP.
Indeed at the 2IKM critical point there are three relevant operators
that correspond to possible symmetry breaking instabilities.
Susceptibilities in the antiferromagnetic, superconducting and
p-h symmetry breaking channels are diverging at the impurity QCP.
Since p-h symmetry is already broken in the lattice system,
only the other two instabilities are left to provide a way to
quench the residual entropy associated to the QCP. In this work
we did not consider the superconducting instability but we
expect it to play a relevant role (and possibly compete with
antiferromagnetism) depending on the specific details of the
band structure of the realistic material.

Notice however several important differences with the picture
of local quantum criticality of Ref.\cite{Si01}. The reference
system describing the physics of the lattice model requires a
cluster of two sites, rather than one, embedded in a medium.
The critical point of this two impurity model, has two relevant
directions corresponding to varying the hybridization and breaking
p-h symmetry. So the impurity model without the self-consistency
condition has a multicritical point rather than a critical point.
Generic problems, as the one considered in this paper, are p-h
asymmetric, and therefore in these cases the two impurity QCP is
not reachable.
However the analysis of the NRG low-energy spectrum confirms that in going
from one side to the other of the crossover (correspondingly, from
one side to the other of the transition in the CDMFT case) the
system evolves from a state in which the impurities are screened
by the conduction electrons (for large $V$) to a state in which
the impurities take advantage of the RKKY interaction to form a
singlet and decouple from the conduction band (for small $V$).
Hence the low-energy picture for $V<V^*$ is that of a conduction
band asymptotically decoupled from a collection of spins arranged
in singlets.
In the lattice model the self-consistency condition transforms
the crossover of the impurity model into a true phase transition.

\begin{figure}
  \begin{center}
    \includegraphics[width=8cm]{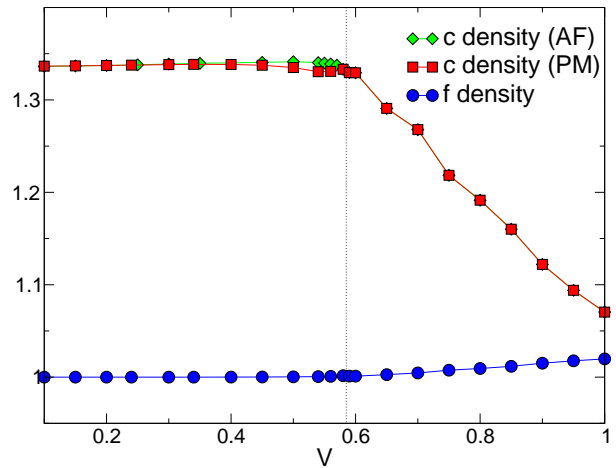}
    \caption{(Color online) $f$ and $c$ electron occupation
    per site. The transition is marked by the vertical dashed
    line.
    \label{densities}}
  \end{center}
\end{figure}

Following the paramagnetic solution we can also observe a
violation of the Luttinger theorem for $V<V^*$ in the form of a
mismatch between the number of particles and the volume of the
Fermi surface. This is clear from the plot of the number of $c$
and $f$ electrons per site as a function of $V$ (Fig.
\ref{densities}). In the PM phase both the conduction and the $f$
bands are doped ($n_f, n_c \neq 1$). In this case the Fermi
surface has a volume equal to the doping $\delta = n_c + n_f - 2$.
The total occupation is not constant because we fixed the chemical
potential rather than the density. For $V<V^*$ the occupation in
both bands becomes essentially constant, implying that $V$ is no
more effective. In this regime the only contribution to the Fermi
surface comes from the $c$ electrons which, being non-interacting
and decoupled from the $f$ band, have a Fermi surface with volume
$n_c = 1 + \delta$.
These results once again support the orbitally
selective Mott transition picture of the KB
\cite{demedici05a,pepin-2006}.
Namely, an $f$-like fluid that becomes incompressible beyond the
KB point.

\begin{figure}
  \begin{center}
    \includegraphics[width=8.5cm]{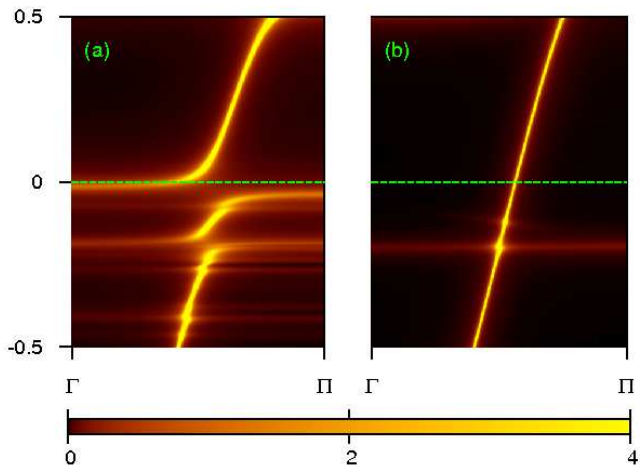}
    \caption{(Color online) Total spectral function as a function
    of momentum and energy on the path connecting the origin
    $\Gamma=(0,0,0)$ and the corner $\Pi=(\pi,\pi,\pi)$ of
    the Brillouin zone. (a) $V=0.65$ PM phase. (b) $V=0.30$
    PM solution in the AF phase.
    \label{bands}}
  \end{center}
\end{figure}

In Fig.\ref{bands} we plot the $k$-resolved total spectral density
$-\frac{1}{\pi} \rm{Im} (G_c(\vect{k},\omega) +
G_f(\vect{k},\omega))$ at low energies along the paramagnetic
solution in the two phases. We intentionally consider points far
from the transition to show that the results are robust and do not
depend on the vicinity to special points in the phase diagram. To
obtain this quantity we reconstruct the rotational and
translational lattice symmetry, broken by the cluster-tiling procedure, by
periodization of the cumulants \cite{Stanescu06b}. This scheme is
more appropriate when correlations are stronger. Indeed the simple
periodization of the self-energy does not reproduce the local
quantities correctly upon integration over momenta, introducing
spurious states in gapped regions of the spectrum. In panel (a)
the system is in the PM phase ($V=0.65$). It is possible to track
the origin of the small value of $Z$ in the narrow band crossing
the chemical potential. The Fermi surface encloses a ``small''
region of the Brillouin zone centered around the origin.
In panel (b) we show the PM solution in the AF phase ($V=0.30$).
It is clear that the volume of the Fermi surface in this phase has
increased. The difference between the Fermi surface volume and the
density is equal to one within the numerical precision, confirming
the violation of the Luttinger theorem as expected in the
orbital selective Mott
transition \cite{pepin-2006}. It is also possible to track the
non-dispersive $f$-band at around $\omega \sim -0.2$.

\begin{figure}
  \begin{center}
    \includegraphics[width=8.0cm]{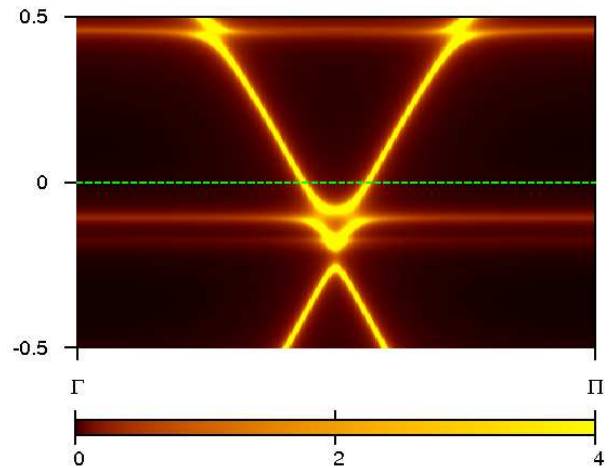}
    \caption{(Color online) Total spectral function as a function
    of momentum and energy along the same path of Fig. \ref{bands}
    for the AF solution. ($V=0.50$)
    \label{bands_AF}}
  \end{center}
\end{figure}

Since the PM solution for $V<V^*$ is unstable to
antiferromagnetism, to make contacts with experiments we plot in
Fig.\ref{bands_AF} the spectral density for the AF solution
($V=0.50$). The bands are doubled because of the AF ordering.
These results are representative for the whole AF phase down to
small value of $V$. The striking difference with respect to the PM
phase is the big rearrangement of the spectral weight: now the
occupied region is centered around the $(\frac{\pi}{2},
\frac{\pi}{2}, \frac{\pi}{2})$ point. This result implies a sudden
jump across the antiferromagnetic transition in all the properties
connected with the topology of the Fermi surface and it is
consistent with the experimental observation of a discontinuous
behavior of the Hall coefficient\cite{Paschen04} and of the de
Haas-van Alphen frequencies\cite{Shishido05}. The contribution of
the $f$ electrons in this phase is manifest in the two
non-dispersing bands at $\omega \sim -0.1$ and $\omega \sim 0.45$,
corresponding to the two peaks in Fig.\ref{spectral_function},
while the bands crossing the chemical potential have mostly $c$
character.

\section{Conclusions \label{conclusions}}

To summarize, we have constructed and investigated a cluster
dynamical mean field theory for the periodic Anderson model using
a link as a reference frame to capture the competition of the
Kondo effect and the RKKY interaction. The mean field phase
diagram away from half-filling features a transition separating a
paramagnetic heavy electron phase from an antiferromagnetic
phase. In the PM phase, the phase transition is accompanied by a
strong renormalization of quasiparticles, marked by a large
enhancement of the effective mass on all the Fermi surface.

By constraining a mean-field metastable paramagnetic solution
into the AF region of the phase diagram, we have been able to show
that the the transition can be
described as an orbitally selective Mott transition.
The decoupling of the $f$-band from the
conduction band marks a localization of the $f$-electrons, as
shown by the opening of a pseudogap in the single particle spectrum
and the appearance of poles in the self-energy at the Fermi
level, a clear fingerprint of ``Mottness''. This is further
supported by the incompressibility of the system, which presents
a constant density by varying the hybridization, and by the violation
of the Luttinger theorem for $V<V^*$ along the PM solution.
It is important to compare this transition, which presents a
orbital-selective Mott character, with the traditional DMFT Mott
transition of the single band Hubbard model. In the present case
the weight is shifted from the chemical potential to the immediate
vicinity forming a pseudogap rather than being transferred to the
Hubbard bands at much higher energy. This feature is the product
of our link-DMFT analysis in three dimensions, strikingly
different from the infinite dimensional DMFT result where the
physics is purely local. It would be important in future DMFT
investigations to better understand the nature of this Mott-like
transition by studying its cluster-size dependence.

In the SDW scenario, the fermions which are not coupled by the
ordering vector are unaffected by the transition. On the contrary
in our CDMFT study, the transition is accompanied by a
dramatic rearrangement of the Fermi surface, in agreement
with recent experiments\cite{Paschen04,Shishido05}.

Our findings can be understood in terms of proximity to a
decoupling QCP with the same properties of the QCP found in the
2IKM, namely the impurity model underlying the self-consistency
equations. Although the lattice system cannot reach that critical point due
to the particle-hole symmetry breaking, the region in which the
effect of the critical point is relevant is large enough to trigger
the lattice antiferromagnetic instability\cite{DeLeo04f}. If we follow the
paramagnetic solution the vicinity to the critical point results
instead in an increase of the non-local correlations $\Sigma_{12}$ in
agreement with the behavior of the 2IKM. In real systems this effect is
preempted by the lattice antiferromagnetic instability.
It would be interesting to realize heavy fermion systems in highly
frustrated lattices to suppress the antiferromagnetic instability.
This would provide a physical realization of the paramagnetic
solution of the CDMFT equations.

In this framework the antiferromagnetic transition is the result
of the response of the system to the instability of the nearby
decoupling QCP, rather than being an intrinsic feature. Notice
that other correlations are diverging at the 2IKM critical point.
We expect that depending on the details of the model other
instabilities (mainly superconductivity) compete with
antiferromagnetism to order the system and quench the residual
entropy of the unstable QCP.

We believe that our description captures the correct physics of
the periodic Anderson model in an intermediate asymptotic
regime of energies and proximity
to the transition.
At higher energies, the single site DMFT description is regained.
At very low energies and very close to the transition, further
non local effects could turn out to be very important.
We do not know if our treatment fits within the local quantum
criticality paradigm where no spatial anomalous dimensions are
generated. To elucidate these questions would require the
investigation of the dynamical susceptibility beyond the cluster
DMFT approximation, which is beyond the scope of this work.

We would like to thank M. Fabrizio, A. Georges, F. Steglich, P. Coleman, C.
Pepin, I. Paul, O. Parcollet, P. Sun, K. Haule, E. Lebanon, M.
Capone, T. Stanescu, J.C. Domenge and L. de' Medici for helpful
discussions and comments. This work was supported by the NSF under
Grant No. DMR 0528969.

\end{document}